# Orbital Hybridization-Induced Ising-Type Superconductivity in a Confined Gallium Layer

Hemian Yi[1,2,8], Yunzhe Liu[1,8], Chengye Dong[3,8], Yiheng Yang[4,8], Zi-Jie Yan[1], Zihao Wang[1], Lingjie Zhou[1], Dingsong Wu[4], Houke Chen[4], Stephen Paolini[1], Bing Xia[1], Bomin Zhang[1], Xiaoda Liu[1], Hongtao Rong[1], Annie G. Wang[1], Saswata Mandal[1], Kaijie Yang[1], Benjamin N. Katz[1], Lunhui Hu[5], Jieyi Liu[4,6], Tien-Lin Lee[6], Vincent H. Crespi[1], Yuanxi Wang[7], Yulin Chen[4], Joshua A. Robinson[1,3], Chao-Xing Liu[1], and Cui-Zu Chang[1]

[1]Department of Physics, The Pennsylvania State University, University Park, PA 16802, USA

[2]Tsung-Dao Lee Institute and School of Physics and Astronomy, Shanghai Jiao Tong University, Shanghai 201210, China

[3]Department of Materials Science and Engineering and Two-dimensional Crystal Consortium, The Pennsylvania State University, University Park, PA 16802, USA

[4]Department of Physics, Clarendon Laboratory, University of Oxford, Parks Road, Oxford OX1 3PU, UK

[5]Center for Correlated Matter and School of Physics, Zhejiang University, Hangzhou 310058, China

[6]Diamond Light Source, Harwell Science and Innovation Campus, Didcot OX11 0DE, UK

[7]Department of Physics, University of North Texas, Denton, TX 76205, USA

[8]These authors contributed equally: Hemian Yi, Yunzhe Liu, Chengye Dong, and Yiheng Yang

Corresponding authors: hemianyi@sjtu.edu.cn (H.Y.); cxl56@psu.edu (C.-X. L.); cxc955@psu.edu (C.-Z. C.).

**Abstract: In low-dimensional superconductors, the interplay between quantum confinement and interfacial hybridization effects can reshape Cooper pair wavefunctions and induce**

novel forms of unconventional superconductivity. In this work, we employ a plasma-free, carbon buffer layer-assisted confinement epitaxy method to synthesize trilayer gallium (Ga) sandwiched between a graphene layer and a 6H-SiC(0001) substrate, forming an air-stable graphene/trilayer Ga/SiC heterostructure. In this confined light element Ga layer, we demonstrate interfacial Ising-type superconductivity driven by atomic orbital hybridization between the Ga layer and the SiC substrate. Electrical transport measurements reveal that the in-plane upper critical magnetic field $\mu_0 H_{c2,\parallel}$ reaches ~21.98 T at $T$ =400 mK, approximately 3.38 times the Pauli paramagnetic limit (~6.51 T). Angle-resolved photoemission spectroscopy (ARPES) measurements combined with theoretical calculations confirm the presence of split Fermi surfaces with Ising-type spin textures at the K and K' valleys of the confined Ga layer strongly hybridized with SiC. Moreover, by incorporating finite relaxation time induced by impurity scattering into an Ising-type superconductivity model, we reproduce the entire temperature-dependent $\mu_0 H_{c2,\parallel}$ phase diagram. This work establishes a new strategy to realize unconventional pairing wavefunctions by combining quantum confinement and interfacial hybridization effects in superconducting thin films. It also opens new avenues for designing scalable superconducting quantum electronic and spintronic devices through interfacial engineering.

**Main text:** The interplay between dimensional confinement and electronic correlations in low-dimensional superconductors has emerged as a key frontier in condensed matter physics, offering a pathway to realizing unconventional superconductivity[1-4]. In conventional superconductors, inversion symmetry and time reversal symmetry ensure energy degeneracy between electrons with momenta $\mathbf{k}$ and $-\mathbf{k}$ for both spin states, leading to $s$-wave spin-singlet pairing as described by the Bardeen-Cooper-Schrieffer (BCS) theory. In contrast, unconventional superconductivity with unique Cooper pairing symmetries stems from various mediating mechanisms, including strong electron correlations[5,6], magnetic fluctuations[7,8], and spin-orbit coupling (SOC) (Refs.[9-13]). Among these, SOC plays a pivotal role in noncentrosymmetric superconductors. The absence of inversion symmetry induces spin splitting, leading to mixed-parity pairing states that combine spin-singlet and spin-triplet components[14]. This SOC effect underpins several exotic pairing types, including Rashba-type[13], Ising-type[9-12], and topological superconductivity[15-17]. Artificially engineering unconventional superconductivity and tailoring Cooper pairing wavefunctions via SOC offer significant potential for developing fault-tolerant topological quantum computations[18,19] and energy-efficient superconducting electronics and spintronics[20].

In three-dimensional superconductors, an external magnetic field suppresses superconductivity through the orbital effect and the Zeeman interaction. However, as superconductors approach the two-dimensional (2D) limit, the in-plane orbital effect is quenched, and Cooper pair-breaking is, instead, governed by the Pauli paramagnetic limit[21]. In a specific class of 2D superconductors, inversion symmetry breaking, combined with SOC, induces Zeeman-like spin-split energy bands near K and K' in the Brillouin zone. This effect aligns electron spins along the out-of-plane direction at K and K' valleys, leading to spin-valley locking[22]. The absence of conventional SU(2) spin rotation symmetry gives rise to Ising-type superconductivity[9,10], characterized by a violation of the Pauli paramagnetic limit under an in-plane magnetic field $\mu_0 H_\parallel$. This behavior has been demonstrated in thin films of transition metal dichalcogenides (e.g., $MoS_2$ and $NbSe_2$) (Refs.[9,10]) and inversion-symmetric Sn(111) films (i.e., stanene)[11,12], both of which rely on spin splitting driven by strong SOC at K (K') and Γ valleys, respectively. It is commonly believed that Ising-type superconductivity typically appears in 2D superconductors composed of heavy elements. Light-element superconductors, such as ultrathin gallium (Ga) layers, are unlikely to host unconventional Cooper pairing due to their weak SOC and spin-degenerate ground states in their freestanding form.

In this work, we synthesize a trilayer Ga sandwiched between an epitaxial graphene layer and a 6H-SiC(0001) substrate to form an air-stable graphene/trilayer Ga/SiC heterostructure through plasma-free and carbon buffer layer-assisted confinement epitaxy (Fig. 1a and Methods). The graphene/trilayer Ga/SiC sandwiches exhibit zero resistance superconducting temperature $T_{c,0}$ ~3.50 K, and a large in-plane upper critical magnetic field $\mu_0 H_{c2,\parallel}$ ~21.98 T at $T$ =400 mK. The $\mu_0 H_{c2,\parallel}$ value is ~3.38 times larger than the Pauli paramagnetic limit, which is estimated as $\mu_0 H_p$ ~1.86$T_{c,0}$ ~6.51 T. Our ARPES measurements reveal a pair of hole Fermi surfaces near K and K' valleys, in good agreement with band structure calculations based on our theoretical model. These concentric Fermi surfaces exhibit Ising-type spin splitting with a magnitude on the order of atomic SOC strength, which originates from atomic orbital hybridization between the bottom Ga layer and the Si layer of the SiC substrate. We find that a quarter circle can fit the normalized $\mu_0 H_{c2,\parallel}$ well as a function of temperature. This unique behavior is interpreted in a theoretical model described by Ising-type pairing wavefunctions intertwined with the Zeeman effect and impurity scatterings. Our results demonstrate orbital hybridization-induced Ising-type superconductivity in graphene/trilayer Ga/SiC, as well as provide insight into interfacial control of inversion-symmetry breaking phases in the artificial heterostructures of light element superconductors.

The confined trilayer Ga films are synthesized on a partially epitaxial graphene-terminated SiC substrate via a plasma-free confinement epitaxy technique. Our samples are composed of ~90% bilayer graphene/trilayer Ga/SiC regions and ~10% monolayer graphene/trilayer Ga/SiC regions (Fig. S1). Note that the top graphene layer protects trilayer Ga from oxidation, making graphene/trilayer Ga/SiC samples air-stable for ex situ characterization and measurements. X-ray photoelectron spectroscopy (XPS) revealed carbon 1*s* peaks of partially epitaxial graphene at binding energies of 284.70 eV and 285.30 eV (Extended Data Fig. 1). These peaks disappear upon Ga intercalation, indicating that the carbon buffer layer has decoupled from the SiC substrate. In addition, the Si-C peak shifts from 283.43 eV to 282.35 eV due to the existence of Ga above SiC. These results are consistent with our prior studies on plasma-assisted Ga intercalation in epitaxial graphene/SiC (Ref.[23]). Cross-sectional scanning transmission electron microscopy (STEM) images show a trilayer Ga between bilayer graphene and the SiC substrate (Figs. 1b, 1c, and Extended Data Fig. 2).

To verify the high crystallinity of the confined trilayer Ga film at the macroscopic scale, the graphene/trilayer Ga/SiC sandwich is examined by X-ray standing wave measurements (Fig. 1d and Extended Data Fig. 3), a well-established technique for high-resolution structural analysis of epitaxial monolayers of 2D materials on single crystal substrates[24-26]. In Fig. 1d, the Ga 3*s* peak, excited by the SiC(0006) standing wave field, exhibits a strong intensity modulation versus photon energy. The characterized coherent fraction $f_{0006}$ ~0.87 is comparable to that of the SiC substrate, confirming a sharp atomic distribution of the Ga element along the *c*-axis. The ultra-low frequency modes at Raman shifts of 25 cm$^{-1}$ and 53 cm$^{-1}$ further confirm the formation of a metallic Ga layer between the graphene and SiC layers (Extended Data Fig. 4). The uniformity of the confined Ga layer is characterized by Raman spectroscopy map of the peak at 25 cm$^{-1}$ (Extended Data Fig. 4d). The homogeneous signals of the Ga layer are observed over a scale of tens of micrometers. Compared to our prior growth method[23], the carbon buffer layer-assisted Ga intercalation prevents the formation of plasma-induced defects. It preserves the structural integrity of the graphene layer, yielding an atomically flat pristine trilayer Ga.

Next, we perform electrical transport and scanning tunneling microscopy/spectroscopy

(STM/S) measurements to confirm the uniform superconductivity in our graphene/trilayer Ga/SiC sandwiches. We have used Samples S1 to S4 for electrical transport measurements. Figure 1e shows $T$ dependence of the sheet longitudinal resistance $R$ for Sample S1. We observe a kink feature at $T$ ~256 K, indicating a first-order phase transition associated with the liquid metal structure of β-phase Ga (Refs.[27,28]). The superconducting phase transition temperature $T_c$, at which $R$ drops to 50% of its normal state value, is ~4.08 K. Figure 1f shows the atomic resolution STM image of a graphene/trilayer Ga/SiC sandwich. We observe a hexagonal lattice structure of the top graphene layer. Through measuring differential conductance $dI/dV$, a uniform superconducting gap with coherence peaks at ±0.22 meV is observed at $T$ =310 mK (Figs. 1g and 1h), consistent with its $T_c$ ~4.08 K (Fig. 1e).

**Pauli-limit violation in graphene/trilayer Ga/SiC**

Next, we investigate the property of the superconducting gap under an out-of-plane magnetic field $\mu_0 H_\perp$. Unlike the molecular beam epitaxy (MBE)-grown bilayer Ga films on a GaN(0001) substrate[29], which exhibit an upper critical magnetic field $\mu_0 H_{c2,\perp}$ of ~800 mT, our graphene/trilayer Ga/SiC sandwiches show a much lower $\mu_0 H_{c2,\perp}$ of ~100 mT (Fig. S2). Moreover, as $T$ increases, the superconducting gap gradually vanishes at $T_c$ ~3 K (Extended Data Fig. 5). We note that $T_c$ determined from STS measurements is slightly lower than that obtained from electrical transport measurements (Fig. 2), presumably due to the proximity-superconductivity on the top surface of the graphene layer in STS measurements.

Figure 2a shows $R$-$T$ curves of Sample S2 under varying $\mu_0 H_\perp$. At $\mu_0 H_\perp$ =0 T, $T_{c,0}$ ~3.50 K, slightly lower than that of bulk β-phase Ga but significantly higher than that of bulk α-phase Ga (Refs.[30,31]). Unlike bulk β-phase Ga, the superconductivity in graphene/trilayer Ga/SiC exhibits much higher sensitivity to small $\mu_0 H_\perp$ (Sample S2 in Figs. 2a and 2c, and Sample S3 in Fig. S3) but remains robust under exceptionally strong $\mu_0 H_\parallel$ (Figs. 2b and 2d). Both $\mu_0 H_{c2,\perp}$ - $T_c$ and $\mu_0 H_{c2,\parallel}$ - $T_c$ phase diagrams are extracted and plotted in Figs. 2e and 2f, respectively. Here the values of $\mu_0 H_{c2,\perp}$ and $\mu_0 H_{c2,\parallel}$ are determined as the magnetic field at which $R$ drops to 50% of the normal state resistance. Note that the $\mu_0 H_{c2,\parallel} \sim T$ curve near $T_c$ fits well with the 2D Ginzburg-Landau (G-L) theory (Fig. 2f), demonstrating the 2D nature of superconductivity in the trilayer Ga film. The coherence length $\xi$ is estimated as $\sim 68.6$ nm by fitting $\mu_0 H_{c2,\perp} \sim T_c$ curve to the 2D G-L theory. The extrapolated $\mu_0 H_{c2,\parallel}$ reaches $\sim 27.89$ T at $T = 0$ K, which exceeds $\mu_0 H_p$ by a factor of $\sim 4.28$. The superconducting thickness $d_{sc}$ is estimated to be $\sim 0.60$ nm, comparable to the thickness of the trilayer Ga ($\sim 0.72$ nm).

**Split Fermi surface with dominant Ising-type spin textures**

To explore the electronic origin of robust superconductivity under $\mu_0 H_\parallel$, we perform ARPES measurements on our graphene/trilayer Ga/SiC heterostructure (Fig. 3). Figure 3a shows the Fermi surface map, where we observe a $\sim 30°$ crystallographic rotation of the Ga layer relative to the graphene layer (Fig. S4). Strong spectral intensity at the Fermi surface is observed near the $K_{Gr}$ points of the graphene layer, corresponding to its Dirac cones. In addition, two small concentric Fermi pockets, denoted as $\beta_1$ and $\beta_2$ in Fig. 3a, are observed near the K valley of the Ga layer. Along the $\Gamma-K$ direction, a larger, nearly-free-electron Fermi pocket $\alpha$ is observed, which splits into $\alpha_1$ and $\alpha_2$ branches (Fig. 3a). Between $\alpha$ and $\beta$ pockets, an additional Fermi pocket $\gamma$ is also observed (Figs. 3a and 3d). Compared to our prior study[23], the observed Fermi surface of the graphene/trilayer Ga/SiC sandwiches synthesized through plasma-free, carbon buffer layer-assisted confinement epitaxy reveals finer band structures and new Fermi surface pockets near the K and K' valleys of the Ga layer. To investigate the evolution of the $\beta_1$ and $\beta_2$ pockets, we analyze constant energy contours at binding energies $E_b = 0$ eV (i.e., the Fermi surface), 0.08 eV, and 0.15 eV (Fig. 3b). As $E_b$ increases, the $\beta_1$ and $\beta_2$ pockets expand and separate, confirming their hole-like character. Although Ga is a light element ($Z$=31), and typically exhibits weak SOC, the unique structure of our graphene/trilayer Ga/SiC sandwiches, specifically the interface between the trilayer Ga and the Si-terminated SiC substrate, can potentially switch on spin splitting through SOC due to quantum confinement and interfacial effects.

To understand the ARPES electronic band structure, we perform first-principles calculations on trilayer Ga/SiC. Among the three most energetically favorable configurations of trilayer Ga, presuming a 1:1 lattice match with SiC, none exactly reproduces the band structure observed in ARPES, particularly the Fermi pockets near K (Extended Data Fig. 6a). This discrepancy suggests that the trilayer Ga/SiC structure cannot be modeled as lattice-matched trilayer Ga based on the following two observations in our cross-sectional STEM measurements. (*i*) The layer spacing between the topmost and middle Ga layers (~0.26 nm) is significantly larger than that between the middle and bottom Ga layers (~0.22 nm). (*ii*) The topmost Ga layer is less resolved than the bottom two Ga layers, implying that the topmost Ga layer has weaker interaction with the bottom two Ga layers and may deviate from their in-plane alignment (Extended Data Fig. 7). To model the decoupling between the topmost Ga layer and the bottom two Ga layers, we calculate the evolution of the band structures as the topmost Ga layer is gradually lifted (Fig. S6). Though the layer decoupling is artificially enforced, the calculated band structure based on a combination of a bilayer Ga/SiC heterostructure and a freestanding Ga monolayer captures the key features of the Fermi pockets near K and K'. However, the band structure in the decoupling limit possesses an additional Fermi pocket near Γ (Fig. S6), which is not observed in ARPES. Therefore, we conclude that achieving the electronic structure in ARPES requires the bands of the topmost Ga layer and the bottom two Ga layers to be strongly coupled near Γ and almost decoupled near K and K'. We discuss the possible microscopic origin of this physical scenario in Supplementary Information.

Our prior studies have suggested a strong electron-phonon interaction for the Fermi pockets near K and K' but not those near Γ(Ref.[23]). To describe the energy bands near K and K', we construct an effective low-energy model that integrates the topmost Ga layer as a decoupled component with a bilayer Ga/SiC heterostructure. The details of this model can be found in Extended Data Fig. 6b and Supplementary Information. A detailed comparison of Fermi pockets near K between ARPES experiments and this theoretical model is shown in Figs. 3b and 3c, which capture the main features of all Fermi pockets α, β, and γ. Among them, the α and β Fermi pockets are mainly from the bilayer Ga on SiC, while the γ Fermi pocket originates from the freestanding monolayer Ga. Moreover, both $\beta_1$ and $\beta_2$ hole pockets increase their sizes when lowering energy in Fig. 3c, consistent with ARPES measurements in Fig. 3b.

Besides the Fermi surface map, we further measure the corresponding band dispersions of the split Fermi pockets. Figure 3d shows the band spectra along the high-symmetry Γ-K-M direction. The $\alpha_1$, $\alpha_2$, and γ bands have negative Fermi velocities along the Γ-K direction, whereas the $\beta_1$ and $\beta_2$ bands show positive Fermi velocities. A pronounced hybridization gap appears at ($E$-$E_F$) ~ -0.8 eV, where electronic bands with opposite Fermi velocities intersect near K. Our ARPES band spectra near K are qualitatively consistent with the calculated band spectra of a freestanding monolayer Ga and a bilayer Ga/SiC heterostructure (Fig. 3e). We find an energy splitting of ~100 meV between $\beta_1$ and $\beta_2$ band spectra around the Fermi momentum $k_F$, which is slightly larger than the first principles calculated energy splitting range from 56 meV to 69 meV. Model calculations reveal that $\beta_1$ and $\beta_2$ hole pockets originate from the bilayer Ga/SiC exhibiting the spin texture characterized by a predominant out-of-plane spin polarization ($S_z$) in opposite directions (Fig. 3f and Extended Data Fig. 8), accompanied by a minimal in-plane Rashba-type spin texture ($S_x$ and $S_y$). The opposite out-of-plane spin orientations near K and K' are consistent with an Ising-type spin-valley locking (Fig. 3g). We note that since inversion symmetry is preserved in a freestanding monolayer Ga, no spin splitting occurs in the topmost monolayer Ga in our calculations.

**Atomic orbital hybridization induced Ising-type superconductivity**

To elucidate the role of SOC in superconductivity, we analyze the $\mu_0 H_{c2,\parallel}$ as a function of $T$ (Fig. 4a). At $T$ =400 mK, the experimentally measured $\mu_0 H_{c2,\parallel}$ reaches ~21.98 T, slightly lower than ~27.89 T obtained from the 2D G-L fit as discussed above. The highly anisotropic feature of superconductivity is underscored by the huge ratio $\mu_0 H_{c2,\parallel}/\mu_0 H_{c2,\perp}$ ~220, corroborating the strong 2D character of the superconductivity in the confined trilayer Ga. Moreover, the $\mu_0 H_{c2,\parallel}$ -$T$ curve deviates significantly from the 2D G-L theory at $T \leq 0.8T_c$. Interestingly, the normalized $\mu_0 H_{c2,\parallel}/\mu_0 H_{c2,\parallel}(T$=400 mK) as a function of $T/T_c$ can be well fitted by an equation $[\mu_0 H_{c2,\parallel}/\mu_0 H_{c2,\parallel}(T$=400 mK)$]^2 + (T/T_c)^2$=1. This deviation from the 2D G-L theory indicates the unconventional nature of superconductivity in our graphene/trilayer Ga/SiC sandwiches.

To account for the large $\mu_0 H_{c2,\parallel}$ and the $\mu_0 H_{c2,\parallel}$ -$T$ phase diagram, possible physical scenarios include the Fulde-Ferrell-Larkin-Ovchinnikov (FFLO) phase[32,33], the Klemm-Luther-Beasley (KLB) model with SOC impurities[34], and Ising- or Rashba-type superconductivity[3,14]. For the FFLO phase observed in organic superconductors and heavy fermion superconductors[35,36], the $\mu_0 H_{c2,\parallel}$ -$T$ curve usually exhibits an upturn feature at $T \leq 0.5T_c$. Such behavior is absent in our graphene/trilayer Ga/SiC sandwiches, suggesting that the Pauli-limit violation is unlikely from an FFLO phase. Instead, our theoretical studies and ARPES measurements point to the dominant Ising-type and subdominant Rashba-type SOC for the hole Fermi pockets at K and K' valleys. SOC can be analogous to impurity scattering, affecting the upper critical magnetic field as described by the KLB model. However, the KLB model fails to fit the entire $\mu_0 H_{c2,\parallel}/\mu_0 H_P$ - $T/T_c$ curve ([Fig. 4a](#)), indicating that the spin-orbit scattering alone cannot account for our observations. As SOC can lift spin degeneracy of band structures and strongly modify the Cooper pairing symmetries, its physical role in large $\mu_0 H_{c2,\parallel}$ will be further explored below.

In graphene/trilayer Ga/SiC sandwiches, broken inversion symmetry induces the dominant Ising-type spin splitting similar to thin films of transition metal dichalcogenides ([Fig. 3g](#)) and sub-dominant Rashba-type spin splitting for the Fermi pockets at K and K' valleys. To investigate the impact of SOC, we calculate $\mu_0 H_{c2,\parallel}$ using three different models: Ising-type SOC, Rashba-type SOC, and a combination of both ([Fig. 4b](#) and [Extended Data Fig. 10](#)). However, none of these models reproduces the $\mu_0 H_{c2,\parallel}/\mu_0 H_P$ - $T/T_c$ curves, even worse than the KLB model. The KLB model incorporates both SOC-induced and spin-independent impurity scattering, which motivates us to introduce disorder broadening, characterized by a finite relaxation time $\tau$, into our SOC model. Using a model with Ising-type SOC and impurity scatterings ([Eq. 9](#) in [Methods](#)), we can fit the $\mu_0 H_{c2,\parallel}/\mu_0 H_P$ - $T/T_c$ curve well with an Ising-type spin splitting $2\Delta_1 = 56$meV and a disorder broadening of $1/\tau \approx 18.1$ meV ([Fig. 4c](#)). As expected, $\mu_0 H_{c2,\parallel}$ decreases as $1/\tau$ increases, consistent with the suppression of superconductivity by impurity scattering. We note that this fit is not unique. Including Rashba-type SOC together with Ising-type SOC and disorder broadening can also yield a good fit to the $\mu_0 H_{c2,\parallel}/\mu_0 H_P$ - $T/T_c$ curve ([Extended Data Fig. 10](#)). However, Rashba-type SOC alone, even combined with disorder broadening, fails to fit the $\mu_0 H_{c2,\parallel}/\mu_0 H_P$ - $T/T_c$ curve.

Next, we discuss the possible sources of disorder broadening in our graphene/trilayer Ga/SiC sandwiches. For instance, ARPES measurements of the Fermi surface reveal interlayer electron scattering between the trilayer Ga and the graphene layer, as confirmed by folded electron pockets from Ga along the $\Gamma$-K direction, coupled via a graphene reciprocal vector ($k =2.86$ Å$^{-1}$) ([Figs. S7](#) to [S9](#)). This interlayer coupling likely contributes to the disorder broadening in graphene/trilayer

Ga/SiC sandwiches. In addition, prior studies[23] have shown that superconductivity in graphene/trilayer Ga/SiC arises from electron-phonon interactions involving small hole pockets near K and K', accompanied by strong intervalley scattering between the K and K' valleys[37]. These scattering processes may also play a role in disorder-induced scattering.

Given success in interpreting the observed $\mu_0 H_{c2,\parallel}/\mu_0 H_P \sim T/T_c$ curves, we next analyze the microscopic origin of strong Ising-type SOC in a bilayer Ga/SiC heterostructure. Near the K valley, the Ising-type spin splitting of the $\beta_1$ and $\beta_2$ Fermi surfaces arises predominantly from the $p_+ = p_x + ip_y$ orbital of the bottom Ga layer. To elucidate the emergence of these ground states, we begin with a freestanding bilayer Ga without SOC. The $p_+$ orbital of the bottom Ga layer (i.e., the bottom Ga layer in a trilayer Ga) and the $p_- = p_x - ip_y$ orbital of the upper Ga layer (i.e., the middle Ga layer in a trilayer Ga) are degenerate, belonging to the 2D irreducible representation $K_3$ of the little group -3'm' at the K point (Supplementary Information). Together with two spin states, denoted as ↑ and ↓ for spin-up and spin-down, we obtain four-fold degenerate states (Fig. 4d). Upon including SOC, these degenerate states split into two doubly degenerate energy states. However, neither Ising- nor Rashba-type spin splitting emerges due to the inversion-time (*PT*) symmetry (Fig. 4d). This scenario changes when inversion symmetry *P* is broken. This can be achieved by incorporating an interface hopping term arising from the orbital hybridization between the bottom Ga layer and the adhered Si layer of SiC. The interfacial coupling leads to a large energy splitting between the $p_+$ orbital of the bottom Ga layer and the $p_-$ orbital of the middle Ga layer ($E_{HYB}$ in Fig. 4e). With SOC, this *PT* symmetry-breaking interlayer coupling lifts the residual double degeneracy, enabling an Ising-type spin polarization near K. We note that the energy scale of inversion breaking is determined by atomic hybridization between the bottom Ga layer and the SiC substrate, with an energy splitting of ~1 eV, which is approximately one order of magnitude larger than the atomic SOC strength of Ga (Refs. [38,39]). Consequently, the Ising spin splitting reaches a magnitude on the order of tens of meV. This Ising spin splitting is governed by $E_{SOC}$, rather than being constrained by the energy scale associated with inversion symmetry breaking (Fig. S13). These results suggest that orbital hybridization across the Ga/SiC interface is the underlying mechanism for strong Ising-type SOC in the graphene/trilayer Ga/SiC sandwiches.

To summarize, we use plasma-free, carbon buffer layer-assisted confinement epitaxy to synthesize a confined Ga trilayer between a graphene layer and a SiC substrate. By leveraging quantum confinement and atomic hybridization, we achieve unconventional superconductivity in this confined light element Ga layer by modulating its Ising-type SOC. Our ARPES measurements combined with theoretical calculations reveal split Fermi surfaces and an Ising-type spin texture. The normalized $\mu_0 H_{c2,\parallel}/\mu_0 H_{c2,\parallel}$( $T$=400 mK) - $T/T_c$ curve follows a unique quarter circle behavior, which can be attributed to the interplay of dominant Ising-type SOC, Zeeman effect, and disorder broadening in graphene/trilayer Ga/SiC. Our work establishes a new strategy for engineering unconventional Cooper pairing wavefunctions by incorporating quantum confinement and interfacial effects in a light-element superconductor. The demonstrated interfacial Ising-type superconductivity offers a promising platform for the development of wafer-scale superconducting quantum electronic and spintronic devices.

**Methods**

**Synthesis of graphene/trilayer Ga/SiC sandwiches**

The semi-insulating 6H-SiC(0001) substrates are chemically cleaned and treated to form a surface with ~90% monolayer graphene/buffer regions and ~10% buffer-only regions. The treated SiC substrates are placed ~2.5 mm above the Ga source and then annealed at ~800 °C for 30 minutes in pure argon gas within a tube furnace (Thermo Fisher Scientific). The argon pressure is maintained at ~500 Torr in the tube furnace, with a flow rate of ~ 50 cm$^3$/min. Ga atoms penetrate the buffer regions and diffuse between the buffer layer and the SiC substrate. A sample structure is formed with ~90% bilayer graphene/trilayer Ga/SiC and ~10% monolayer graphene/trilayer Ga/SiC. Finally, these sandwich samples are naturally cooled to room temperature for subsequent ex situ characterization and measurements. The top graphene layer makes these samples air stable.

**Raman spectroscopy measurements**

Raman spectroscopy measurements are conducted using a Horiba LabRAM HR system with a 532 nm laser (~5 mW power) and a 300 grooves mm$^{-1}$ grating. Raman spectra are acquired with a 15 second integration time and normalized to the SiC folded transverse acoustic (FTA) mode at ~150 cm$^{-1}$.

**XPS measurements**

XPS measurements are conducted in a Physical Electronics Versa Probe II equipped with a monochromatic Al Kα X-ray source ($hv$ ~ 1486.7 eV) and a concentric hemispherical analyzer. High-resolution XPS spectra are acquired from a ~200 μm analysis area with a pass energy of ~28 eV for C 1$s$ and Ga 3$d$. All XPS spectra are charge-corrected by referencing the C-C peak in the C 1$s$ spectra at ~284.5 eV.

**STEM measurements**

The samples for our STEM measurements are prepared using in situ lift-out through focused ion beam (FIB) milling. To prevent damage and contamination during the FIB process, an amorphous ~400 nm carbon layer is deposited onto the sample surface before milling. High-resolution STEM measurements are performed on an FEI Titan$^3$ G2 operating at the accelerating voltage of ~200 kV, with a probe convergence angle of ~30 mrad, a probe current of ~70 pA, and high-angle annular dark-field (HAADF) detector angles of 51~300 mrad.

**X-ray standing wave measurements**

The X-ray standing wave measurements are conducted in a chamber with a base vacuum better than $2 \times 10^{-10}$ mbar at beamline I09 of Diamond Light Source, UK. In the X-ray standing wave technique, a standing wave field is generated by the interference between an incident X-ray beam and its Bragg-diffracted beam from a crystal substrate[40-42]. This standing wave is parallel to the (*hkil*) Bragg planes and has periodicity equal to the corresponding interplanar spacing $d_{hkil}$. As the photon energy is scanned through the Bragg condition, the standing wave field shifts by $d_{hkil}/2$ along the (*hkil*) direction. This results in modulations of the photoelectron yield from the near-surface region, which reflect the spatial distributions of the emitters along the (*hkil*) over the spacing $d_{hkil}$. These distributions can be quantitatively extracted by analyzing the yield curves.

In our X-ray standing wave measurements, the standing wave field is generated using the (0006) Bragg reflection of the 6H-SiC(0001) substrate in a backscattering geometry. This reflection is excited by tuning a Si(111) double-crystal monochromator to a photon energy of ~2.465 keV. The X-ray beam spot on the sample is ~$350 \times 350$ $\mu m^2$. At each photon energy step, a photoemission spectrum modulated by the standing wave field is recorded by a Scienta Omicron EW4000 electron analyzer aligned with the X-ray polarization direction. Simultaneously, the diffracted beam from the 6H-SiC(0006) reflection is imaged on a fluorescent screen, and its intensity is recorded using a charge-coupled device (CCD) camera. For our X-ray standing wave data analysis, the reflectivity and the photoelectron yield curves are fitted using the dynamical theory of X-ray diffraction[43]. This analysis yields two normalized free parameters: the coherent fraction and the coherent position, which describe the width and the center of the atomic distribution of the emitters, respectively[40-42]. More details about X-ray standing wave measurements are found in Supplementary Information.

**ARPES measurements**

The in-house ARPES measurements are performed at room temperature in a chamber with a base vacuum better than $5 \times 10^{-11}$ mbar. The energy analyzer used in our in-house ARPES measurements is DA30L (ScientaOmicron). The excitation light is a helium-discharged lamp with a photon energy of ~21.2 eV. The energy and angle resolutions are ~10 meV and ~0.1º, respectively. Synchrotron ARPES measurements are conducted at Beamline Bloch, Max IV, Sweden, in a chamber with a base vacuum better than $8 \times 10^{-11}$ mbar. The energy analyzer used in our synchrotron ARPES measurements is DA30L (ScientaOmicron). The energy and angle resolutions are set to ~25 meV and ~0.2º, respectively. For synchrotron ARPES, the beam spot size is ~6 × 25 $\mu m^2$. The samples are annealed at ~300 °C for 2 hours before ARPES measurements to remove surface contaminants and residual moisture.

**Electrical transport measurements**

The graphene/trilayer Ga/SiC sandwiches are scratched into a Hall bar geometry using a computer-controlled probe station. The effective areas of the Hall bar devices used for electrical transport measurements are ~ 1000 × 500 $\mu m^2$ for Sample S1 and ~ 500 × 500 $\mu m^2$ for Samples S2, S3, and S4. The electrical contacts for transport measurements are made by pressing indium spheres on the Hall bar. The electrical transport measurements under low magnetic fields (≤ 14T) are conducted using two Physical Property Measurement Systems (Quantum Design DynaCool 1.7 K, 9 T/14 T) for $T \geq 1.7$ K and a dilution refrigerator (Bluefors, 10 mK, 9-1-1 T) for $T < 1.7$ K. Transport measurements under high magnetic fields (>14 T) are carried out in a hybrid magnet (Cell 6, 41.5 T, He³) at the National High Magnetic Field Laboratory (NHMFL), Tallahassee. Unless otherwise specified, the excitation current is ~1 μA in all electrical transport measurements.

**STM/S measurements**

The STM/S measurements are performed in a Unisoku 1300 system with a base vacuum better than $3 \times 10^{-10}$ mbar. The system incorporates a single shot $^3$He cryostat to achieve a base temperature of ~310 mK. The maximum magnetic field of the system is ~11 T. Polycrystalline PtIr tips are used in all STM/S measurements. Before conducting STM/S measurements on graphene/trilayer Ga/SiC sandwiches, the PtIr tip is regularly conditioned on an MBE-grown Ag film. The $dI/dV$ spectra are obtained using the standard lock-in method by applying an additional small a.c. voltage at a frequency of ~983 Hz. All STM images are processed with WSxM 5.0 software[44].

**First-principles calculations**

All first-principles calculations are performed using the Vienna Simulation Package (VASP)[45,46], where we employ projector augmented wave pseudopotentials[47,48] and the Perdew-Burke-Ernzerhof (PBE) parametrization of the generalized gradient approximation for the exchange-correlation functional[49]. Unless otherwise specified, a plane-wave expansion energy cutoff of 500 eV is used, with a relaxation force threshold of 10 meV/Å, a $k$-point sampling equivalent to a $21 \times 21 \times 1$ grid for a lateral unit cell of 6H-SiC(0001) and the SiC substrates are modeled as 7 formula unit thick slabs. In our first-principles model of Ga/SiC, the graphene overlayer is omitted, as our prior work[23] has shown that its inclusion does not affect the low-energy band structure of the Ga layer and the charge transfer effect is negligible. For models involving finite-size flakes, supercells with lateral dimensions of $10 \times 10$ (SiC lateral unit cells) and 2-formula-unit-thick SiC slabs are used. The finite-size flakes are hexagonal with an initial diameter of 8 lattice constants (SiC lateral unit cells), ensuring sufficient size convergence to minimize edge effects.

To further verify that the contraction of the topmost Ga layer is not due to trivial finite-size effects, in addition to the third Ga layer on bilayer Ga/SiC models discussed, the same test is performed on a finite-size flake of the second layer Ga stacked on monolayer Ga/SiC. Note that the flake only exhibits minimal contraction, within ~2%, thereby delineating that the loss of templating effects starts at the third Ga layer.

**Theoretical modeling**

As noted in the main text, the band structure near K and K' of a trilayer Ga/SiC heterostructure can be modeled as a bilayer Ga/SiC heterostructure with a decoupled freestanding monolayer Ga. Because all relevant Fermi surfaces for superconductivity originate from the bilayer Ga/SiC heterostructure, we next focus on the effective model for bilayer Ga/SiC derived from our first-principles calculations. To construct this model, we employ the maximally localized Wannier function method to develop a tight-binding model based on $p_x$, $p_y$, and $p_z$ orbitals of Ga atoms and $p_z$ orbital of Si atoms. This model can reproduce well the band structure from our first-principles calculations (Extended Data Fig. 6b and Supplementary Information).

Our prior study [23] has shown that the electron-phonon coupling is strong for the Fermi pockets $\beta_1$ and $\beta_2$ near K. Therefore, we focus on the two low-energy bands that form the Fermi pockets $\beta_1$ and $\beta_2$ near the K point of the Fermi surface. In our tight-binding model, these two bands primarily originate from the hybridization between the $p_z$ orbital of the upper Ga layer and the $p_x + ip_y$ orbital of the bottom Ga layer in bilayer Ga/SiC. Our strategy is to construct the effective model from symmetry considerations[50] and then fit it to our first-principles band structures. The wave-vector group at the K point of the bilayer Ga/SiC heterostructure is 3m', generated by a three-fold rotation $C_{3z}$ and $M_x T$, where $M_x$ is the mirror symmetry along the $x$ direction and $T$ is the time reversal symmetry. Based on this symmetry group, the two-band effective model reads

$$H_K(\mathbf{k}) = (\alpha \mathbf{k}^2 - \mu)\sigma_0 + \Delta_1 \sigma_z + b_x \sigma_x + \beta_R(k_x \sigma_y - k_y \sigma_x) \tag{1}$$

on the spin basis with $\sigma_0$ and $\sigma_{x,y,z}$ representing identity and Pauli matrices for spin. In the above effective Hamiltonian, $\alpha k^2$ is the quadratic-$k$ term of kinetic energy, $\mu$ is the Fermi energy, and the Zeeman splitting is given by $b_x \sigma_x$, where $b_x = \mu_0 H_{c2,\parallel}$ and $\mu_0$ is the Bohr magneton. The $\Delta_1 \sigma_z$ term gives the Ising spin splitting and $\beta_R(k_x \tau_y - k_y \tau_x)$ is the Rashba term. $\alpha = 1.74 a_0^2 \text{eV}$, $\mu = -0.12 \text{eV}$, $\Delta_1 = 0.028 \text{eV}$, and $\beta_R = 0.036 a_0 \text{eV}$ with $a_0$ as the lattice constant can produce the Fermi pockets $\beta_1$ and $\beta_2$ near the K point.

The effective model near K' can be related to that at K by time reversal $T$ and is given by

$$H_{K'}(\mathbf{k}) = (\alpha \mathbf{k}^2 - \mu)\sigma_0 - \Delta_1 \sigma_z + b_x \sigma_x + \beta_R(k_x \sigma_y - k_y \sigma_x). \tag{2}$$

For the superconducting state, we consider spin-singlet intervalley pairing between the Fermi pockets at K and K'. To achieve in-plane upper critical magnetic field $\mu_0 H_{c2,\parallel}$, we solve the linearized gap equation[51]:

$$1 = V_0 \chi_{SC} \tag{3}$$

where $V_0$ is the attractive intervalley pairing interaction parameter, and the superconducting susceptibility $\chi_{SC}$ is given by

$$\chi_{SC} = \sum_{\mathbf{k}, \omega_n} Tr(\sigma_y G_e(i\omega_n, \mathbf{k}) \sigma_y G_h(i\omega_n, \mathbf{k})) \tag{4}$$

where $G_e(i\omega_n, \mathbf{k})$ and $G_h(i\omega_n, \mathbf{k})$ are Matsubara Green's functions for electron and hole, respectively, and are defined as

$$G_e(i\omega_n, \mathbf{k}) = (i\omega_n - H_K(\mathbf{k}))^{-1} \tag{5}$$

$$G_h(i\omega_n, \mathbf{k}) = \left(i\omega_n + H_{K'}^*(-\mathbf{k})\right)^{-1} \tag{6}$$

with the Matsubara frequency $\omega_n$ and momentum $\mathbf{k}$. To take into account the impurity scattering, we make the replacement $\omega_n \rightarrow \omega_n(1 + \frac{1}{2\tau|\omega_n|})$ in the Matsubara Green's function[52], where $\tau$ is the scattering time. The impurity scattering renormalizes the imaginary part of Fermi energy by raising the scattering time. The linearized gap equation can be solved as

$$\ln\left[\frac{T}{T_c}\right] = \frac{1}{2}\left[\Phi(\rho_-, \tau) + \Phi(\rho_+, \tau)\right] + \frac{1}{2}\left[\Phi(\rho_-, \tau) - \Phi(\rho_+, \tau)\right]\frac{\Delta_1^2 + 2\Lambda_f^2 - \left(\mu_0 H_{c2,\parallel}\right)^2}{D_{f-} - D_{f+}} +$$

$$\Psi\left(\frac{1}{2} + \frac{1}{4\pi k_b T_c \tau}\right) - \Psi\left(\frac{1}{2}\right) \tag{7}$$

where $\Phi(\rho) = Re\left[\Psi\left(\frac{1}{2}\right) - \Psi\left(\frac{1+i\rho}{2} + \frac{1}{4\pi k_b T \tau}\right)\right]$ with $\Psi$ as the digamma function. Other parameters are defined as $\Lambda_f = \beta_R k_F$ with $k_F$ as Fermi momentum measured from the K point, and $\rho_\pm = \frac{i}{2\pi k_B T}(D_{f+} \pm D_{f-}))$ with $D_{f\pm} = \sqrt{\left(\Lambda_f \pm \mu_0 H_{c2,\parallel}\right)^2 + \Lambda_f^2 + \Delta_1^2} \tag{8}$

In the limit of zero Rashba SOC, $\beta_R = 0$, $\rho_+ = 2\sqrt{\left(\mu_0 H_{c2,\parallel}\right)^2 + \Delta_1^2}$, $\rho_- = 0$, and

$\frac{\Delta_1^2 + 2\Lambda_f^2 - \left(\mu_0 H_{c2,\parallel}\right)^2}{D_{f-} - D_{f+}} = 1 - \frac{2\left(\mu_0 H_{c2,\parallel}\right)^2}{\left(\mu_0 H_{c2,\parallel}\right)^2 + \Delta_1^2}$. Eq. 7 changes to

$$\ln\left[\frac{T}{T_c}\right] = \Phi(\rho_-, \tau) - \left[\Phi(\rho_-, \tau) - \Phi(\rho_+, \tau)\right]\frac{\left(\mu_0 H_{c2,\parallel}\right)^2}{\left(\mu_0 H_{c2,\parallel}\right)^2 + \Delta_1^2} + \Psi\left(\frac{1}{2} + \frac{1}{4\pi k_b T_c \tau}\right) - \Psi\left(\frac{1}{2}\right) = \Psi\left(\frac{1}{2} + \right.$$

$$\left.\frac{1}{4\pi k_b T_c \tau}\right) - \Psi\left(\frac{1}{2} + \frac{1}{4\pi k_b T \tau}\right)\frac{\Delta_1^2}{\Delta_1^2 + \left(\mu_0 H_{c2,\parallel}\right)^2} - Re\Psi\left[\frac{1}{2} + \frac{1}{4\pi k_b T \tau} + i\frac{\sqrt{\Delta_1^2 + \left(\mu_0 H_{c2,\parallel}\right)^2}}{2\pi k_b T}\right]\frac{\left(\mu_0 H_{c2,\parallel}\right)^2}{\Delta_1^2 + \left(\mu_0 H_{c2,\parallel}\right)^2} \tag{9}$$

where $T_c$ is the superconducting temperature at $\mu_0 H$ =0 T, $2\Delta_1$ is the Ising spin splitting energy, and $b_x = \mu_0 H_{c2,\parallel}$ is the energy of an external Zeeman field with $\mu_0$ being the Bohr magneton.

Eq. 9 is the main result of this section and can be used to fit the $\mu_0 H_{c2,\parallel}/\mu_0 H_P$ - $T/T_c$ curve in Fig. 4. By including impurity scattering and Ising SOC, we obtain an excellent fit for the $\mu_0 H_{c2,\parallel}/\mu_0 H_P$ - $T/T_c$ curve (Fig. 4c). Note that including Rashba SOC also yields a good fit for the $\mu_0 H_{c2,\parallel}/\mu_0 H_P$ - $T/T_c$ curve (Extended Data Fig. 10).

**Acknowledgments:** We thank N. Samarth and J. Zhu for helpful discussions and M. Leandersson,

J. Osiecki, and C. Polley for technical assistance. This project is primarily supported by the Penn State MRSEC for Nanoscale Science (DMR-2011839), including sample synthesis, STEM, XPS, Raman, and electrical transport measurements. The STM/S measurements are partially supported by the NSF-CAREER award (DMR-1847811) and the NSF grant (DMR-2241327). The in-house ARPES measurements are performed in the NSF-supported 2DCC MIP facility (DMR-2039351). CZC acknowledges the support from the Gordon and Betty Moore Foundation's EPiQS Initiative (GBMF9063 to C. -Z. C). The work done at the National High Magnetic Field Laboratory is supported by NSF Cooperative Agreement No. DMR-2128556 and the State of Florida. HY acknowledges the Shanghai Pujiang Program (25Z510101786). YC acknowledges Diamond Light Source for time on beamline I09 under proposal No. NT37930. We acknowledge the MAX IV Laboratory for beamtime on the BLOCHs beamline under proposal No. 20230668 and No. 20240633. Research conducted at MAX IV, a Swedish national user facility, is supported by Vetenskapsrådet (Swedish Research Council, VR) under contract 2018-07152, Vinnova (Swedish Governmental Agency for Innovation Systems) under contract 2018-04969 and Formas under contract 2019-02496.

**Author contributions:** HY and CZC conceived and designed the experiment. HY, ZY, LZ, and CZC conducted electrical transport measurements. HY, HR, and CZC performed in-house ARPES measurements. YY, DW, HC, and YC performed synchrotron ARPES measurements. YY, JL, and TL performed X-ray standing wave measurements. CD and JAR synthesized all samples and performed STEM, XPS, and Raman measurements. ZW, SP, BX, and CZC performed STM/S measurements. LZ, BZ, and XL performed electrical transport measurements in a dilution refrigerator. YL, SM, KY, BNK, LH, VHC, YW, and CXL provided theoretical support. HY, YL, YW, CXL, and CZC analyzed the data and wrote the manuscript with input from all authors.

**Competing interests:** The authors declare no competing financial interests.

**Data availability:** The datasets generated during and/or analyzed during this study are available from the corresponding author upon request.

**Figures and figure captions:**

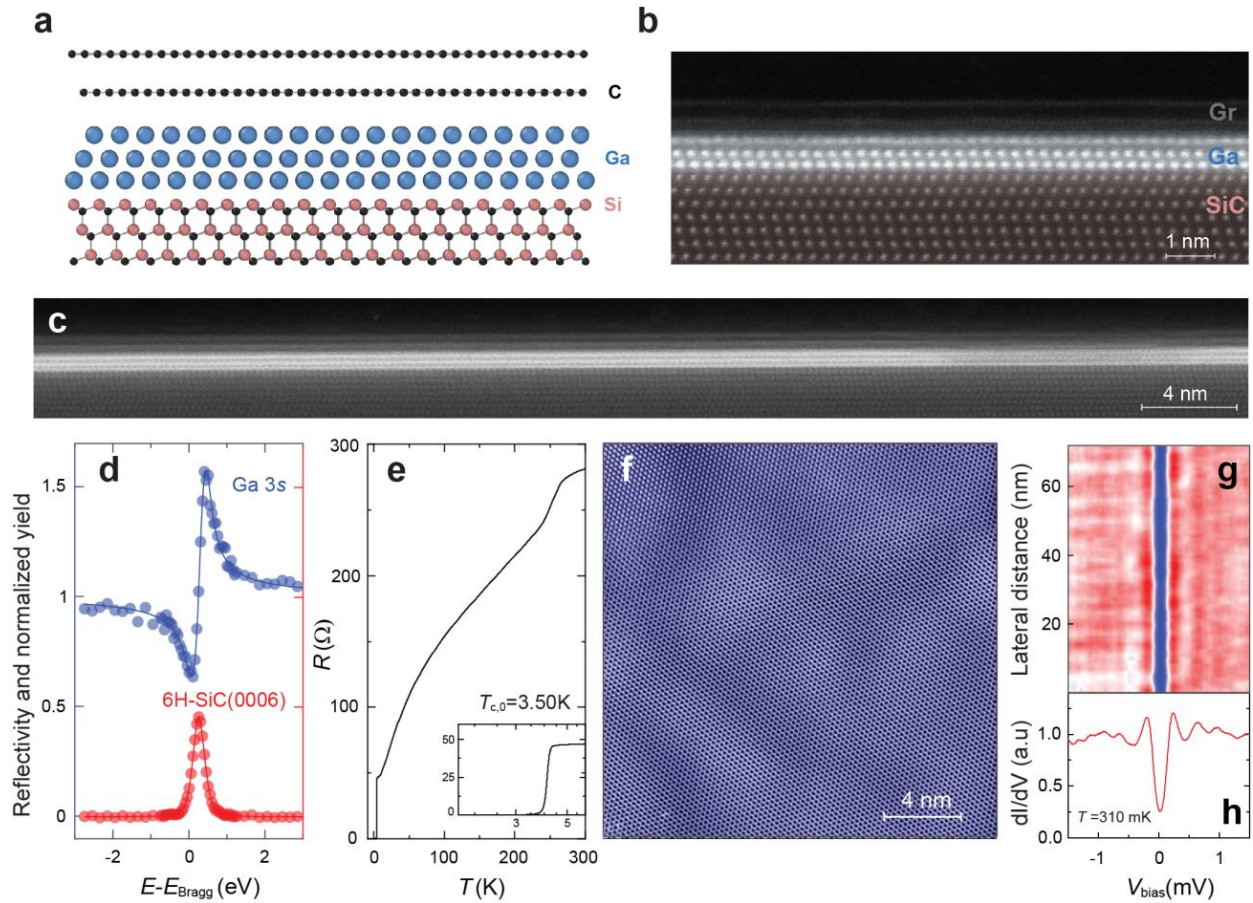

**Fig. 1| Characterization of graphene/trilayer Ga/SiC. a**, Schematic of the confined trilayer Ga.
**b**, **c**, Cross-sectional STEM images. **d**, X-ray standing wave spectra of the Ga layer using the 6H-
SiC(0006) reflection. The blue and red dots correspond to the normalized Ga $3s$ photoelectron
yield and the (0006) Bragg reflectivity of the 6H-SiC(0001) substrate, respectively. For Ga $3s$
peaks, the coherent fraction $f_{0006}$ is ~0.87. **e**, $R$-$T$ curves of Sample S1. Inset: the enlarged $R$-$T$
curve. The $T_{c,0}$ and $T_{c,onset}$ values are ~3.50 K and ~4.23 K, respectively. **f**, Atomic resolution STM
image (sample bias $V_B$ = +50 mV, tunneling current $I_t$ = 500 pA, and $T$ = 310 mK). **g**, $dI/dV$
spectra on the surface of a graphene/trilayer Ga/SiC sandwich ($V_B$ = +1.5 mV, $I_t$ =500 pA, and $T$ =
310 mK). **h**, A representative superconducting gap ($V_B$ = +1.5 mV, $I_t$ =500 pA, and $T$ = 310 mK).

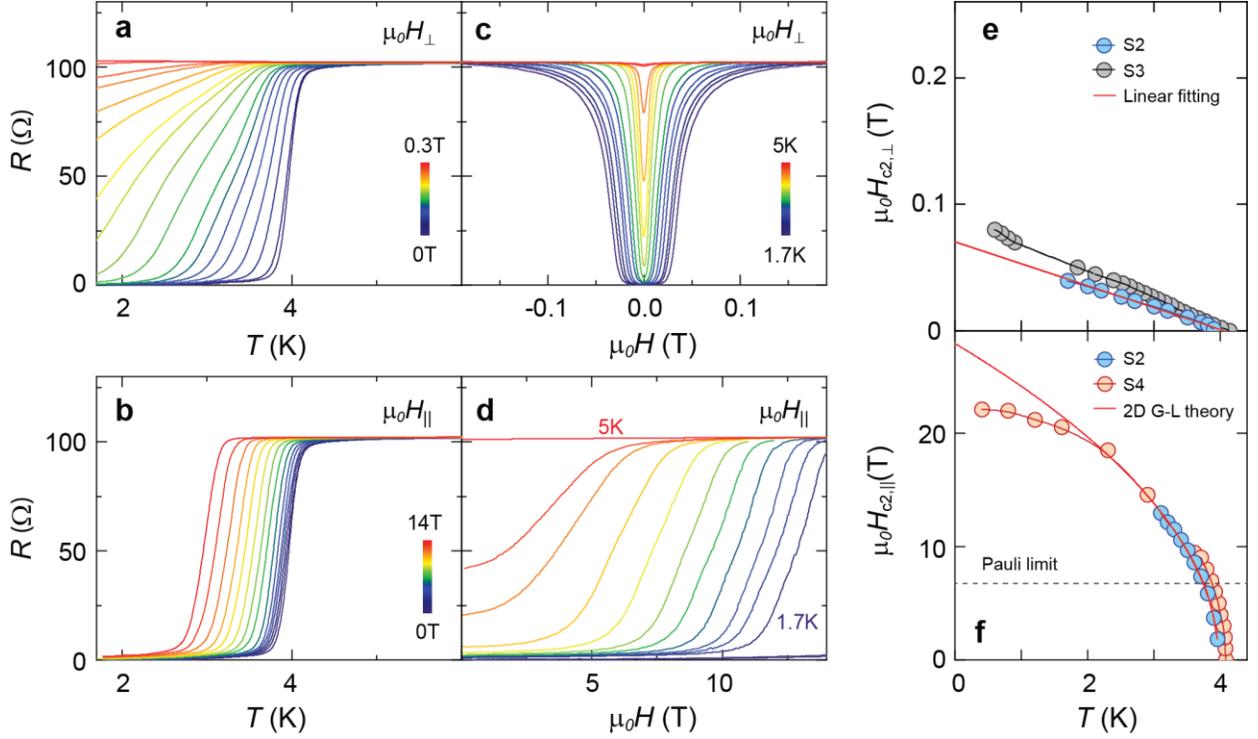

**Fig. 2| Pauli limit violation in graphene/trilayer Ga/SiC. a, b,** $R$-$T$ curves of Sample S2 measured under varying magnetic fields: (**a**) $\mu_0 H_\perp$ and (**b**) $\mu_0 H_\parallel$. **c, d,** $R$-$\mu_0 H$ curves of Sample S2 measured at different temperatures: (**c**) $\mu_0 H_\perp$ and (**d**) $\mu_0 H_\parallel$. **e, f,** Temperature dependence of $\mu_0 H_{c2,\perp}$ (**e**) and $\mu_0 H_{c2,\parallel}$ (**f**) of Samples S2, S3, and S4. All experimental data points are extracted from (**a–d**), Extended Data Fig. 9, and Fig. S3. The red curves in (**e, f**) represent fits using the 2D G–L theory for Sample S2. The dashed line in (**f**) indicates the Pauli paramagnetic limit, $\mu_0 H_p = 1.86 T_{c,0}$ ~6.51 T.

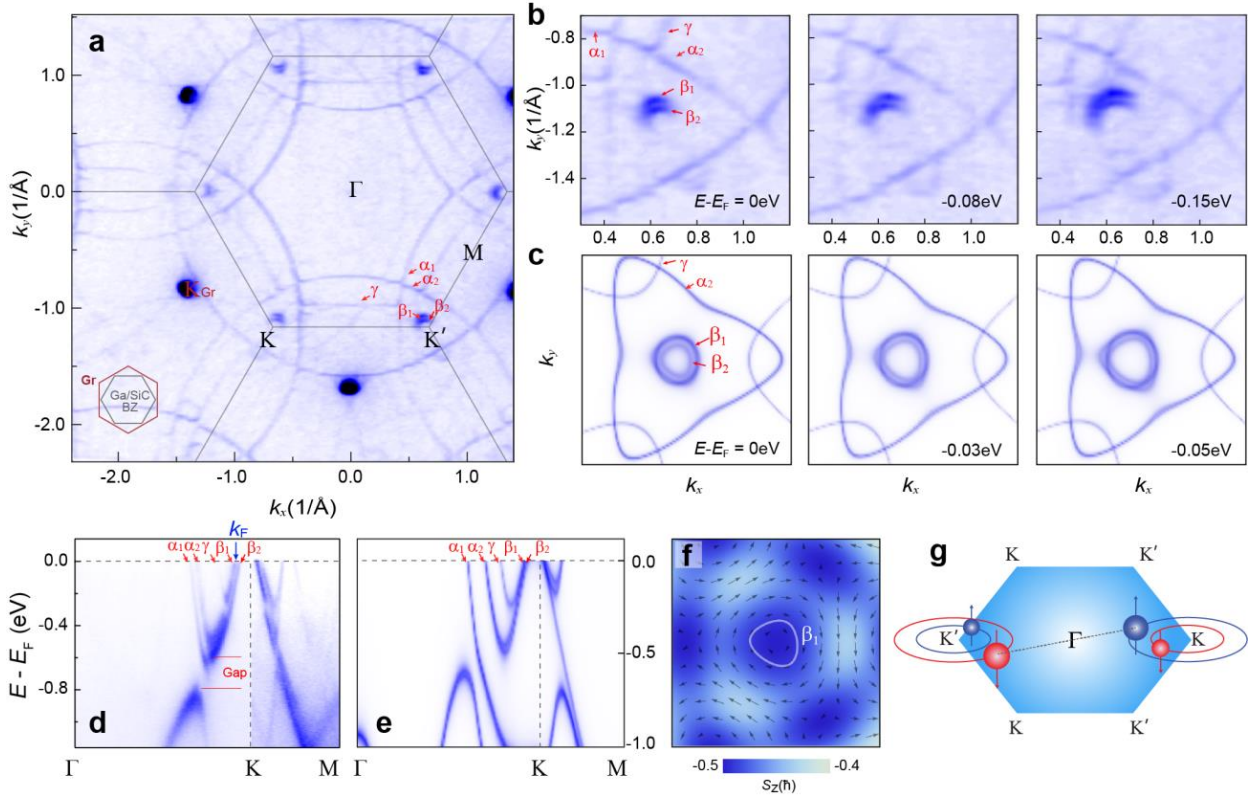

**Fig. 3| Split Fermi surface with Ising-type spin textures in graphene/trilayer Ga/SiC. a,** Fermi surface map. The black and red hexagons denote the Brillouin zones of the trilayer Ga and graphene, respectively. **b, c,** Enlarged constant energy contours (**b**) and calculated constant energy contours (**c**) near the K valley of the trilayer Ga. **d, e,** ARPES band spectra (**d**) and calculated band spectra (**e**) along the Γ-K-M direction. **f,** Spin texture of bilayer Ga with upper spin polarization along the out-of-plane direction. **g,** Schematic of the Fermi surface and spin polarization near the K and K' valleys in the first Brillouin zone of bilayer Ga. The ARPES measurements are performed using photon energy of 150 eV at $T$ =18 K.

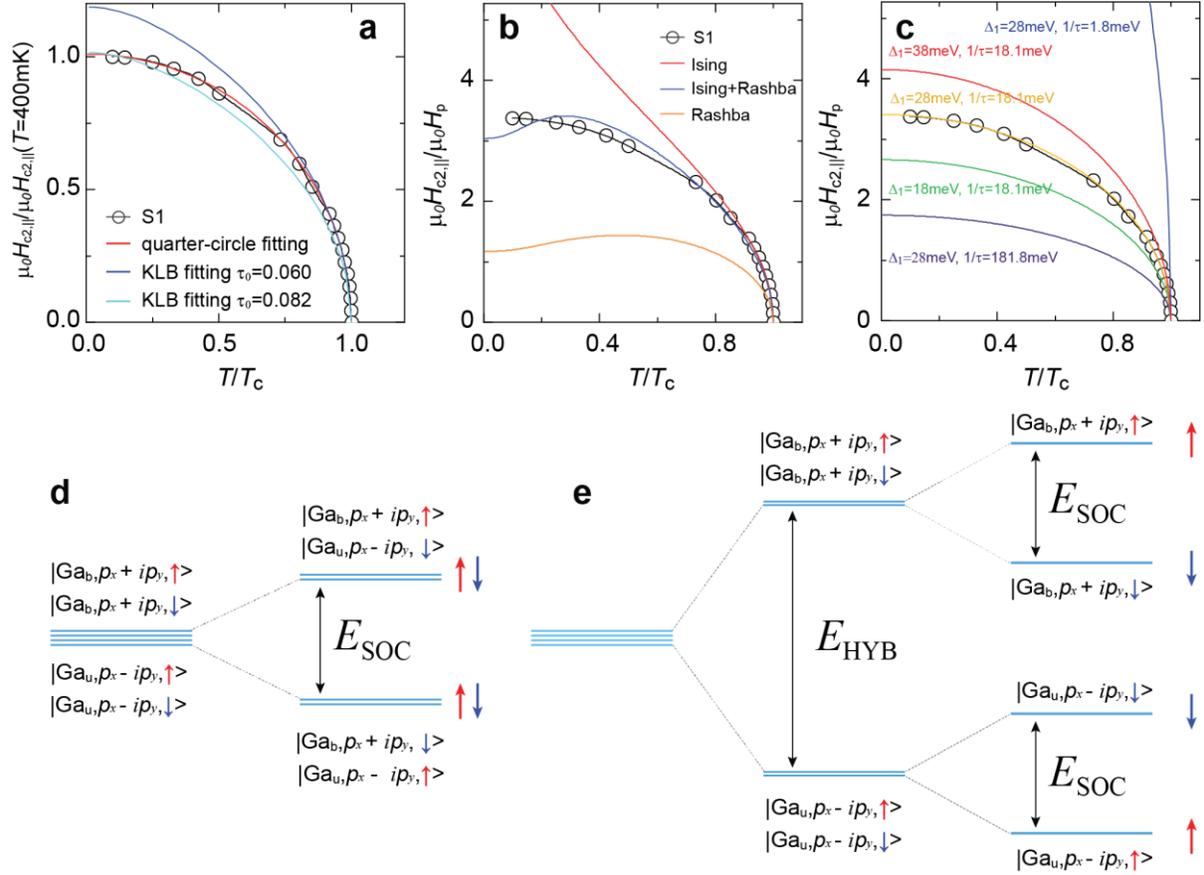

**Fig. 4| Atomic orbital hybridization-induced Ising-type superconductivity in graphene/trilayer Ga/SiC. a,** Reduced temperature $T/T_c$ dependence of normalized in-plane upper critical magnetic field, $\mu_0 H_{c2,\parallel}/\mu_0 H_{c2,\parallel}$ ($T=400$ mK). The red curve represents a quarter-circle fit, while the blue and cyan curves correspond to the KLB fit with two different $\tau_0$ values. **b,** $T/T_c$ dependence of the normalized $\mu_0 H_{c2,\parallel}/\mu_0 H_p$. The red and yellow curves correspond to fits based on Ising- and Rashba-type superconductivity, respectively, while the blue curve represents a fit that incorporates both Ising and Rashba SOC. **c,** Ising-type superconductivity in graphene/trilayer Ga/SiC with varying degrees of disorder. **d,** Schematic of the fourfold-degenerate states in a freestanding bilayer Ga. The energy associated with SOC is denoted as $E_{SOC}$. Ising- or Rashba-type SOC spin splitting is absent due to the preserved inversion symmetry. **e,** Schematic of Ising-type spin splitting in a bilayer Ga/SiC heterostructure. Orbital hybridization lifts the degenerated energy states between Ga layers, breaking inversion symmetry. The hybridization energy is denoted as $E_{HYB}$. Due to the resultant inversion symmetry breaking, the SOC then induces the formation of Ising-type spin-polarized energy states.

**Extended Data Figures**

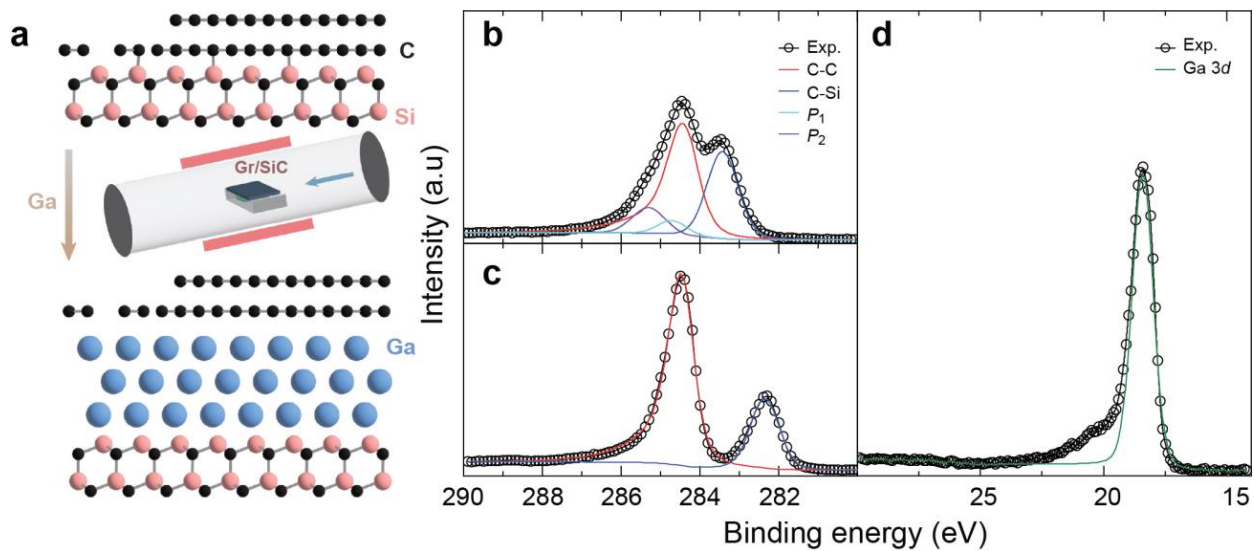

**Extended Data Fig. 1| XPS spectra of graphene/trilayer Ga/SiC. a**, Schematic of the Ga intercalation process in a SiC substrate with a surface of ~90% monolayer graphene/buffer regions and ~10% buffer-only regions. **b, c,** XPS spectra of C 1*s* before (**b**) and after (**c**) the Ga intercalation process. The $P_1$ and $P_2$ components in the C 1*s* spectra of partially epitaxial graphene disappear after the Ga intercalation, indicating decoupling of the buffer layer from the SiC substrate. **d**, XPS spectra of Ga 3*d* after the Ga intercalation process. The monolayer graphene forms when the ~10% buffer-only region decouples from the SiC substrate during Ga intercalation.

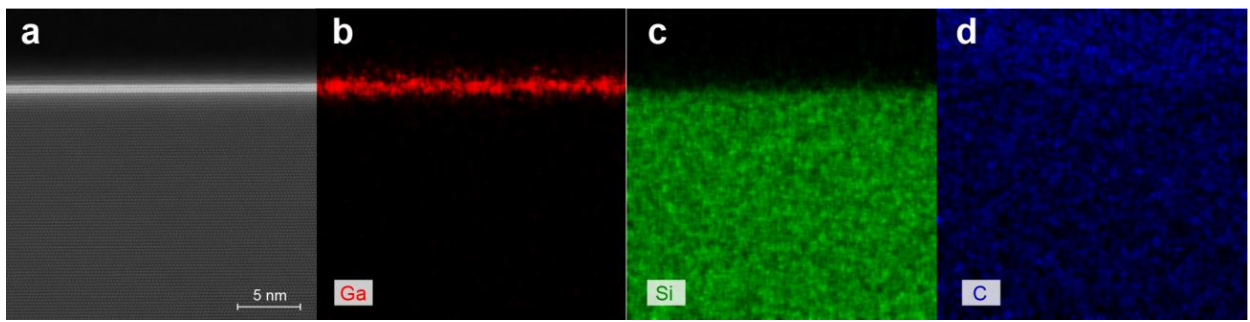

**Extended Data Fig. 2| STEM image and corresponding energy dispersive X-ray spectroscopy (EDS) maps of graphene/trilayer Ga/SiC. a**, STEM image of a graphene/trilayer Ga/SiC heterostructure. **b-d,** The corresponding EDS maps of Ga (**b**), Si (**c**), and C (**d**).

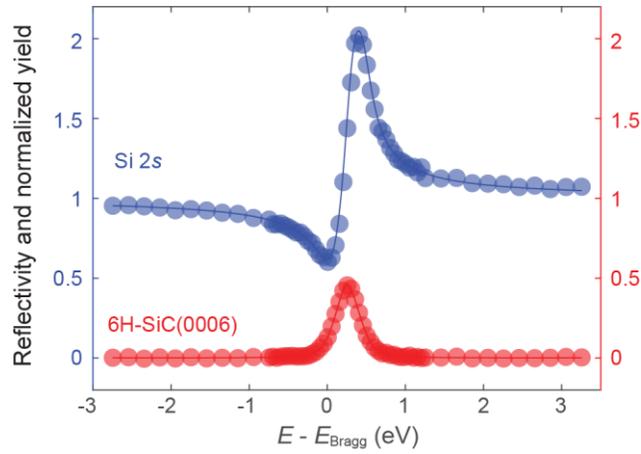

**Extended Data Fig. 3| X-ray standing wave spectra of the Si 2*s* peaks using the 6H-SiC(0006) reflection.** The blue and red dots correspond to the normalized Si 2*s* photoelectron yield and the (0006) Bragg reflectivity of the 6H-SiC(0001) substrate, respectively. Since the SiC lattice is used to generate X-ray standing waves, the coherent fraction $f_{0006}$=1 is expected for the Si 2*s* peak, with a polarization factor $P_{0006}$=1. In our experiment, the measured $f_{0006}$=1.02 and $P_{0006}$=1.01 confirms the reliability of our X-ray standing wave measurements.

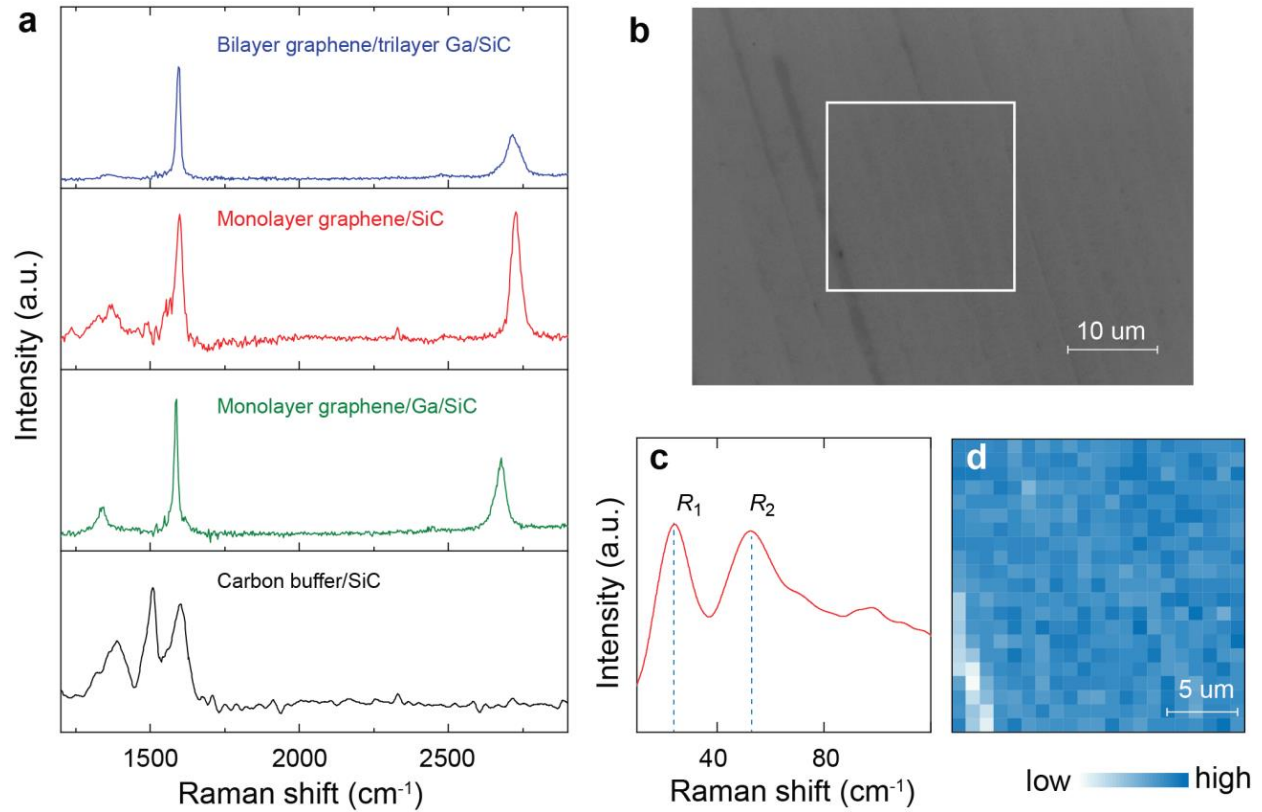

**Extended Data Fig. 4| Raman spectra of graphene/trilayer Ga/SiC. a**, Raman spectra of bilayer graphene/trilayer Ga/SiC, bilayer graphene/SiC, monolayer graphene/trilayer Ga/SiC, and monolayer graphene/SiC, respectively. **b**, Optical microscopy image of a graphene/trilayer Ga/SiC sandwich. **c**, Raman spectra of the graphene/trilayer Ga/SiC sandwiches exhibit two low-frequency metal modes at $25 \, \text{cm}^{-1}$ and $53 \, \text{cm}^{-1}$. The $R_1$ and $R_2$ peaks emerge only after Ga intercalation, confirming their origin in the trilayer Ga layer. **d**, Raman map of the $R_1$ peak at $25 \, \text{cm}^{-1}$ over a $20 \times 20 \, \mu\text{m}^2$ aera, as marked in (**b**).

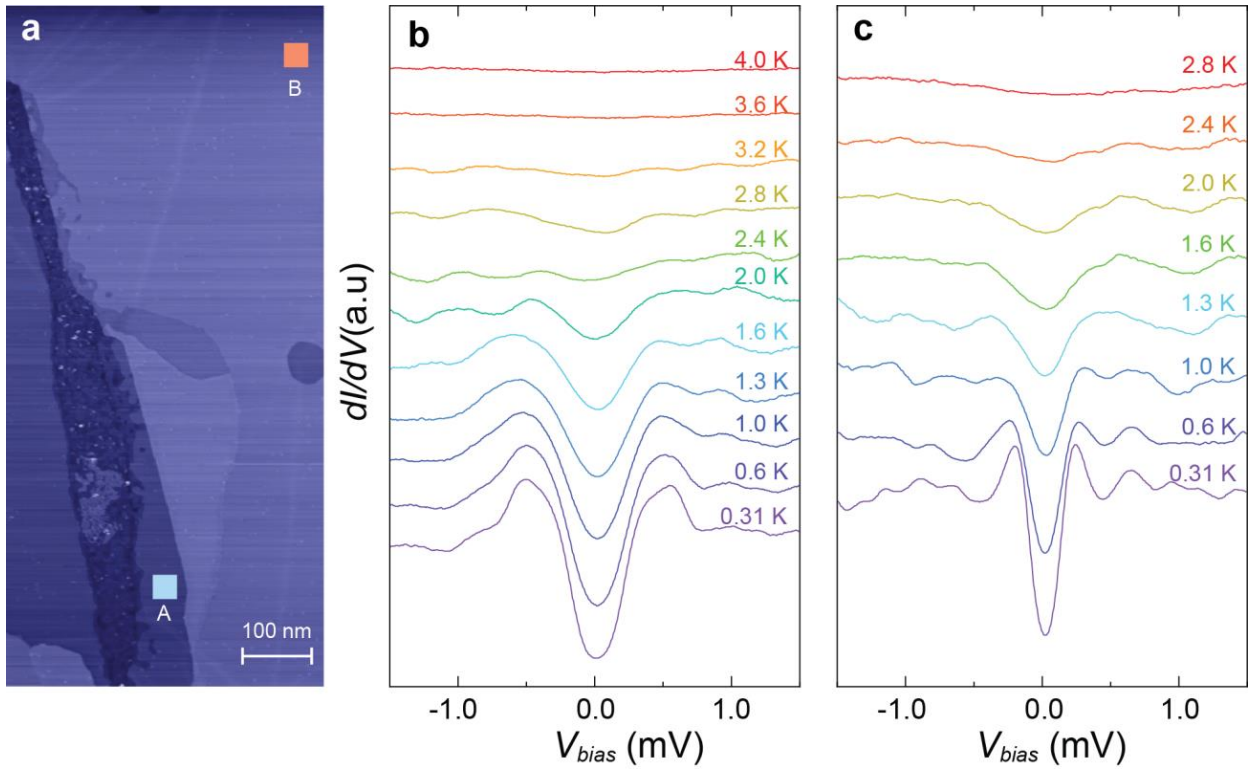

**Extended Data Fig. 5| Temperature-dependent *dI/dV* spectra of graphene/trilayer Ga/SiC. a,** Large-scale STM image of a graphene/trilayer Ga/SiC heterostructure ($V_B$ = +100 mV, $I_t$ = 100 pA, and $T$ = 310 mK). **b, c,** Temperature-dependent *dI/dV* spectra of monolayer graphene/trilayer Ga/SiC (**b**) and bilayer graphene/trilayer Ga/SiC (**c**) heterostructures ($V_B$ = +1.5 mV, $I_t$ =500 pA, and $T$ = 310 mK).

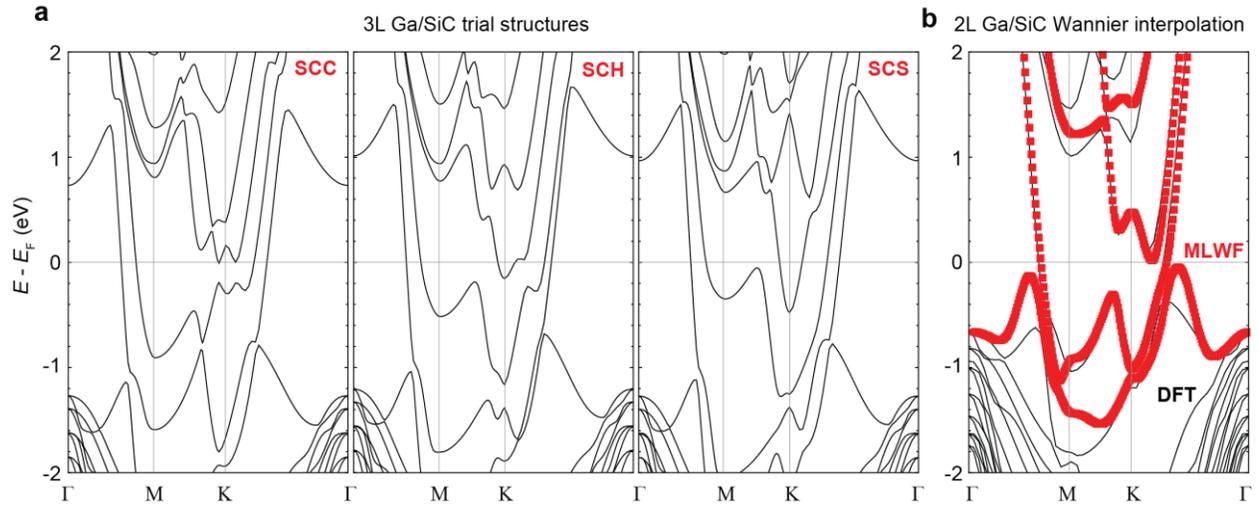

**Extended Data Fig. 6| The calculated band structures of trilayer Ga/SiC and bilayer Ga/SiC heterostructures. a**, Band structures of trilayer Ga/SiC with three different structural orientations. **b**, Band structures of bilayer Ga/SiC. The red squares and the black curves represent the band structures obtained using maximally localized Wannier functions (MLWF) and first-principles calculations, respectively.

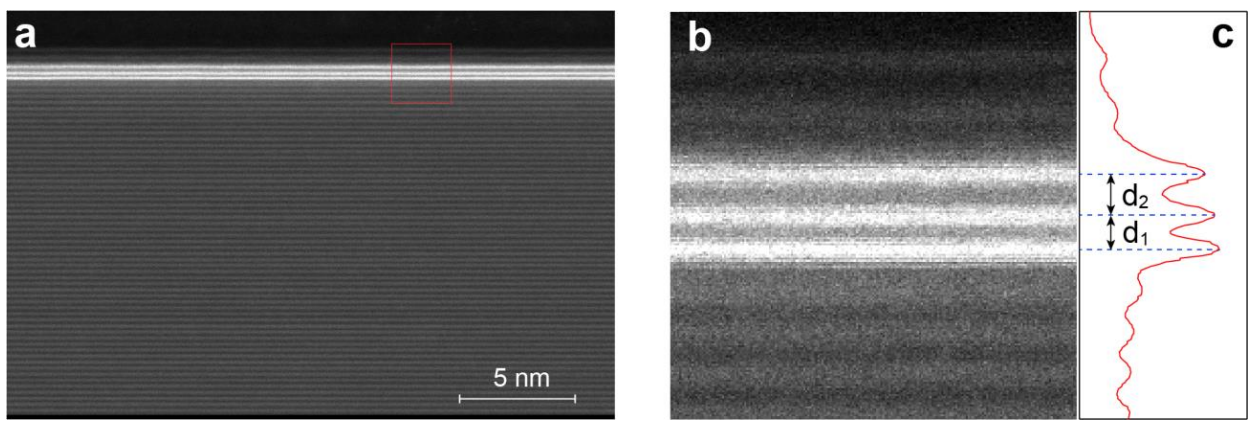

**Extended Data Fig. 7| More STEM images of graphene/trilayer Ga/SiC sandwiches. a,** STEM image of a graphene/trilayer Ga/SiC sandwich. **b,** Enlarged STEM image of the area in (**a**). **c,** Integrated intensity along the perpendicular direction. The spacing between the bottom two Ga layers is $d_1$ =0.22 nm, while the spacing between the top two Ga layers is $d_2$ =0.26 nm. The analysis of Fig.1b gives comparable values, with $d_1$ =0.21 nm and $d_2$ =0.27 nm, respectively.

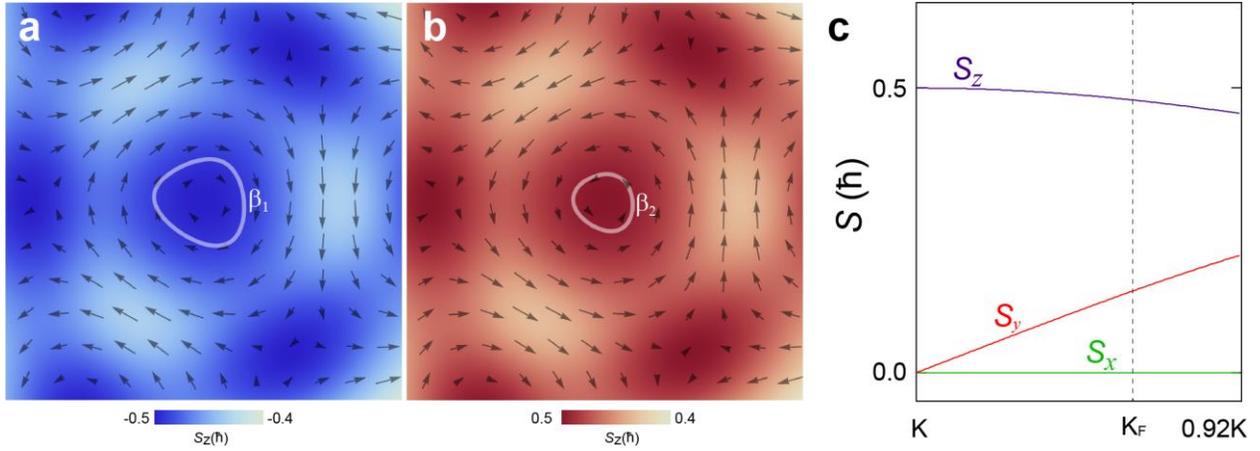

**Extended Data Fig. 8| Spin texture of bilayer Ga near the K valley. a, b,** Spin textures of bilayer Ga showing the $\boldsymbol{S}_z$ component for spin-up (**a**) and spin-down (**b**) states. **c,** Spin polarization with the $S_x$, $S_y$, and $S_z$ components.

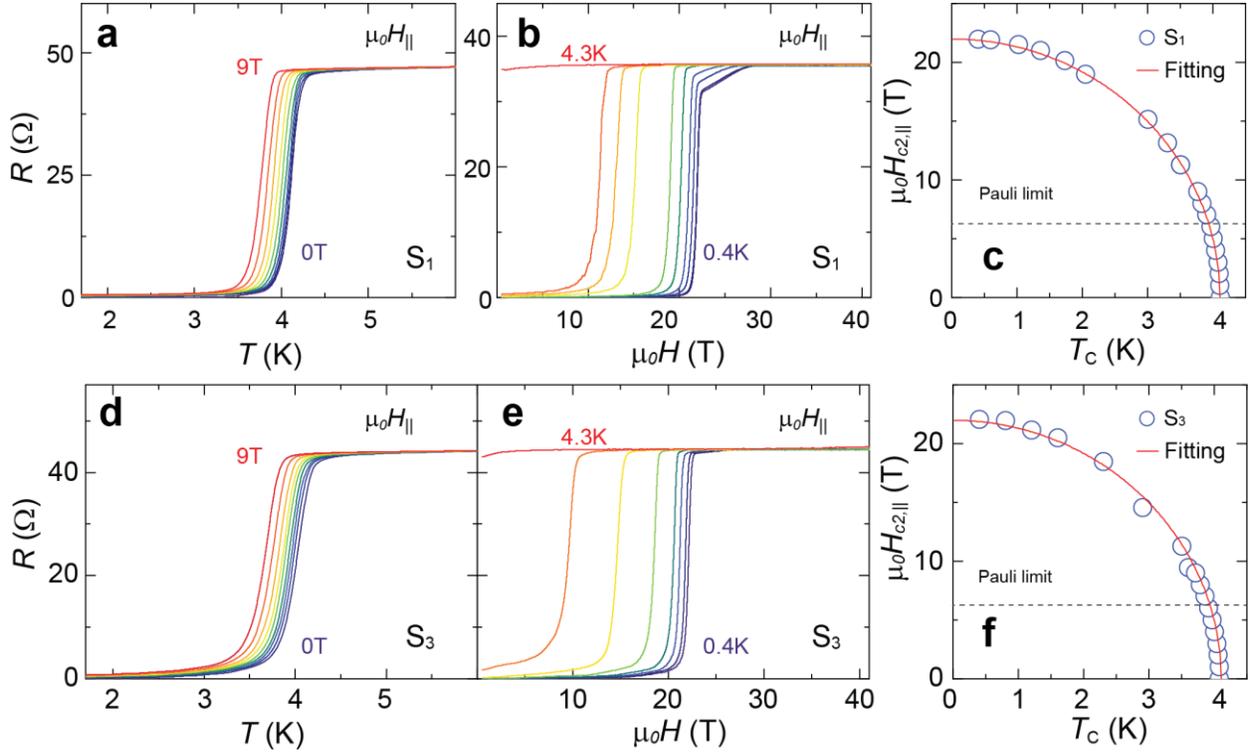

**Extended Data Fig. 9| More transport results of Samples S1 and S3. a,** *R-T* curves of Sample S1 under varying $\mu_0 H_\parallel$. **b,** *R*-$\mu_0 H$ curves of Sample S1 measured under $\mu_0 H_\parallel$ at different temperatures. **c,** Temperature dependence of $\mu_0 H_{c2,\parallel}$. All data points in (**c**) are extracted from (**a**) and (**b**). **d-f,** Same as (**a-c**), but on Sample S3. All data points in (**f**) are extracted from (**d**) and (**e**). The fitting parameters in (**c**) and (**f**) are adopted from those in Fig. 4c ($\Delta_1$ =28 meV and $1/\tau \approx 18.1$ meV).

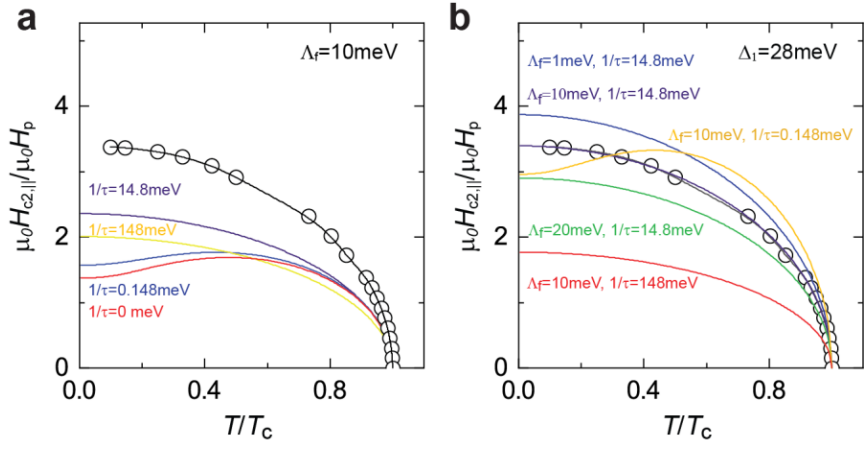

**Extended Data Fig. 10| Theoretical calculations of $\mu_0 H_{c2,\parallel}$ for Sample S1. a,** Calculated $\mu_0 H_{c2,\parallel}/\mu_0 H_p$ - $T/T_c$ with varying impurity scattering strength $1/\tau$. The strength of Rashba- and Ising-type SOC are fixed at $\Lambda_f = 10$ meV and $\Delta_1 = 0$ meV, respectively. **b,** Calculated $\mu_0 H_{c2,\parallel}/\mu_0 H_p$ - $T/T_c$ with $\Delta_1 = 28$ meV. $\Lambda_f$ and $1/\tau$ are varied.